%% file: skeleton.tex
\documentclass[a4paper,11pt]{article}
\usepackage{pos}
\usepackage{deluxetable}
\usepackage{multirow}
\usepackage{pifont}
\definecolor{burgundy}{rgb}{0.5, 0.0, 0.13}
\definecolor{applegreen}{rgb}{0.55, 0.71, 0.0}
\newcommand{\xmark}{\textcolor{burgundy}{\ding{55}}}%
\newcommand{\cmark}{\textcolor{applegreen}{\ding{51}}}%

\title{Target of Opportunity observations of flaring blazars with H.E.S.S. 
}
 \ShortTitle{Target of Opportunity observations of flaring blazars with H.E.S.S. 
}

\author*[a]{M. Cerruti}
\author[b]{C. Boisson}
\author[c]{M. Böttcher}
\author[c]{O. Chibueze}
\author[c]{I. Davids}
\author[c]{A. Dmytriiev}
\author[d]{G. Grolleron}
\author[e]{F. Jankowsky}
\author[d]{J.P. Lenain}
\author[b]{A. Luashvili}
\author[e,c]{M. Zacharias}

\affiliation[a]{Université Paris Cité, CNRS, Astroparticule et Cosmologie, F-75013 Paris, France}

\affiliation[b]{Laboratoire Univers et Théories, Observatoire de Paris, Université Paris Sciences Lettres, CNRS, Université Paris Cité, F-92190 Meudon, France}

\affiliation[c]{Centre for Space Research, North-West University, Potchefstroom 2520, South Africa
}

\affiliation[d]{Laboratoire de Physique Nucléaire et des Hautes Energies (LPNHE), Sorbonne Université, Université Paris Cité, CNRS/IN2P3, F-75005, Paris, France}

\affiliation[e]{Landessternwarte, Universität Heidelberg, Königstuhl, 69117, Heidelberg, Germany}

\onbehalf{for the H.E.S.S. Collaboration} 

\emailAdd{cerruti@apc.in2p3.fr}

\abstract{Blazars are the most common class of TeV extragalactic emitters. In the framework of the AGN unified model, they are understood as AGNs with a relativistic jet pointing close the line of sight. They are characterized by extreme variability, observed to be as fast as minutes. These flares are usually observed at multiple wavelengths and their study require fast reaction and coordination among multiwavelength observatories. An important part of blazars observations with the H.E.S.S. array of Cherenkov telescopes is thus in the form of Target of Opportunity (ToO) observations. In this contribution the H.E.S.S. blazar ToO program is presented, with a focus on recent results. }

\ConferenceLogo{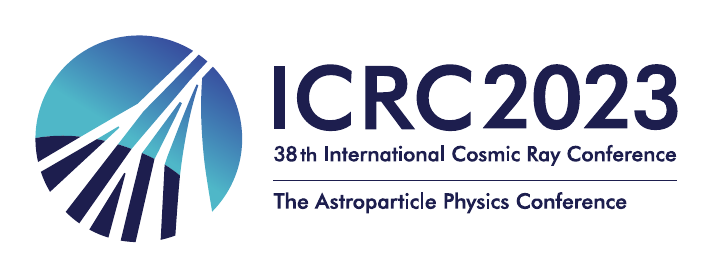}

\FullConference{%
38th International Cosmic Ray Conference (ICRC2023)\\
  26 July - 3 August, 2023\\
  Nagoya, Japan}

\begin{document}
\maketitle

\section{The H.E.S.S. Blazar Target of Opportunity  Program}

The summary of H.E.S.S. Target of Opportunity observations of blazars from 2016 to Spring 2023 is provided in Table \ref{tab:summary}. Columns are: observing season, observing period (defined by full Moons), the observed target, its blazar sub-class, its redshift, the exposure in hours, the type of trigger, if the observations resulted in detection by H.E.S.S., and references.\\

\clearpage
\begin{deluxetable}{c c c c c c c c c}

\tablewidth{0pt}

\tablecaption{Summary table of the H.E.S.S. blazar ToO program from 2016. \label{tab:summary}}

\tablehead{Observing & Observing  & Source & Blazar   & Redshift & Exposure & Trigger & Detection & Ref.\\
Season & Period & & Type & & [h] & & }
\startdata
\hline
\multirow{10}{*}{2016}& P2016-03 & PKS 2022-077 & FSRQ  & 1.388 & 1.6 & LAT & \xmark  & \cite{Emery2023}\\ 
& P2016-05 & PKS 1510-089 & FSRQ  & 0.36 & 16.2 & VHE & \cmark  & \cite{HESS1510}\\ 
& P2016-07 & OT 081 & LBL  & 0.32 & 14.6 & LAT & \cmark  & \cite{OT081}\\ 
& P2016-08 & CTA 102 & FSRQ  & 1.032 & 13.0 & Optical & \xmark  & \cite{HESSToO17}\\ 
& P2016-09 & PKS 0447-439 & HBL  & 0.343 & 6.5 & VHE & \cmark  & \cite{HESSToO17}\\ 
& P2016-10& PKS 2247-131 & ?BL  & 0.22?  & 6.8 & LAT & \xmark  & \\ 
& P2016-11 & PKS 0507+17 & FSRQ  & 0.44 & 0.8 & LAT & \xmark  & \\ 
& P2016-12 & Mrk 421 & HBL  & 0.03 & 2.5 & VHE & \cmark  & \\ 
& P2017-01 & OJ 287 & ?BL  & 0.306 & 2.0 & X-rays & \xmark  & \\ 
& P2017-02 & 3C 279 & FSRQ  & 0.536 & 5.1 & Optical & \xmark  & \cite{Emery3C279}\\ 
\hline
\multirow{5}{*}{2017}& P2017-03 & 3C 279 & FSRQ  & 0.536 & 4.1 & Optical & \xmark  & \cite{Emery3C279}\\ 
& P2017-06 & 3C 279 & FSRQ  & 0.536 & 0.5 & LAT & \xmark  & \cite{Emery3C279}\\ 
& P2017-09/10 & PKS 2022-077 & FSRQ  & 1.388 & 10.5 & LAT & \xmark  & \cite{Emery2023}\\ 
& P2017-12 & Mrk 421 & HBL  & 0.03 & 1.5 & VHE & \cmark  & \\ 
& P2018-01 & 3C 279 & FSRQ  & 0.536 & 8.0 & LAT & \cmark  & \cite{Emery3C279}\\ 
\hline
\multirow{9}{*}{2018}& P2018-02 & 3C 279 & FSRQ  & 0.536 & 4.3 & LAT & \xmark  & \cite{Emery3C279}\\ 
& P2018-03 & TXS 0506+056 & IBL  & 0.337 & 3.8 & LAT & \xmark  & \\ 
& P2018-04 & PKS 0903-57 & FSRQ  & 0.695 & 1.0 & LAT & \xmark  & \\ 
& P2018-04 & M 87 & RG  & 0.004 & 14.7 & VHE & \cmark  & \\ 
& P2018-06 & 3C 279 & FSRQ  & 0.536 & 23.0 & LAT & \cmark  & \cite{Emery3C279}\\ 
& P2018-07 & AP Lib & LBL  & 0.049 & 7.0 & LAT & \cmark  & \\ 
& P2018-09 & PKS 0346-27 & FSRQ  & 0.99 & 2.8 & LAT & \xmark  & \cite{Emery3C279}\\ 
& P2018-11 & PKS 0625-354 & HBL?  & 0.055 & 23.0 & VHE & \cmark  & \cite{PKS0625}\\ 
& P2018-13 & 1ES 1218+304 & HBL  & 0.182 & 11.5 & VHE & \xmark  & \\
\hline
\multirow{5}{*}{2019}& P2019-03/04 & PG 1553+113 & HBL  & $\simeq$ 0.4 & 13.5 & Optical & \cmark  &\\ 
& P2019-07 & PKS 0346-27 & FSRQ  & 0.99 & 5.3 & LAT & \xmark  & \\ 
& P2019-12 & PKS 0208-512 & FSRQ  & 1.003 & 8.5 & LAT & \xmark  & \\ 
& P2020-01 & 3C 273 & FSRQ  & 0.158 & 2.5 & LAT & \xmark  &\\ 
& P2020-02 & PKS 1156-221 & FSRQ  & 0.565 & 1.0 & LAT & \xmark  & \\ 
\hline
\multirow{4}{*}{2020}& P2020-04 & PKS 0903-57 & HBL  & 0.695 & 13.8 & LAT & \cmark  & \cite{Puhlhofer}\\ 
& P2020-06 & PKS 1156-221 & FSRQ  & 0.565 & 8.5 & LAT & \xmark  & \\ 
& P2020-10 & BL Lac & IBL  & 0.069 & 9.8 & LAT & \xmark  & \\ 
& P2020-12 & PKS 0513-459 & FSRQ  & 0.194 & 10.0 & LAT & \xmark  &\\ 
\hline
& & & & & & & & \\
\hline
\multirow{11}{*}{2021}& P2021-02 & PKS 1127-145 & FSRQ  & 1.19 & 9.8 & LAT & \xmark  & \\ 
& P2021-03 & PKS 0837+012 & FSRQ  & 1.12 & 10.0 & LAT & \xmark  & \\ 
& P2021-05 & GB6 J1058+2817 & ?BL  & 0.82 & 1.5 & LAT & \xmark  & \\ 
& P2021-05 & PKS 0027-426 & FSRQ  & 0.492 & 1.3 & LAT & \xmark  &\\ 
& P2021-06 & PKS 1454-354 & FSRQ  & 1.424 & 10.5 & LAT & \xmark  & \\ 
& P2021-06 & PKS 1313-333 & FSRQ  & 1.21 & 9.7 & LAT & \xmark  & \\
& P2021-07 & PKS 1334-127 & FSRQ  & 0.54 & 9.7 & LAT & \xmark  & \\ 
& P2021-07/08 & BL Lac & IBL  & 0.069 & 1.7 & LAT & \xmark  & \\ 
& P2021-09/10 & PKS 0301-721 & FSRQ  & 0.823 & 9.2 & LAT & \xmark  &\\ 
& P21-11->22-01 & PKS 0346-27 & FSRQ  & 0.99 & 31.5 & LAT & \cmark  &  \citep{PKS0346}\\ 
& P2021-13 & PKS 0903-57 & FSRQ  & 0.695 & 1.0 & LAT & \xmark  &\\ 
\hline
\multirow{6}{*}{2022}& P2022-04/05 & PKS 1954-388 & FSRQ  & 0.63 & 36.4 & LAT & \xmark  & \\ 
& P2022-05 & PKS 1127-145 & FSRQ  & 1.19 & 1.9 & LAT & \xmark  & \\ 
& P2022-06 & PKS 0035-252 & FSRQ  & 0.498 & 1.5 & LAT & \xmark  & \\ 
& P2022-07 & PMN J1717-5155 & FSRQ  & 1.16 & 7.5 & LAT & \xmark  &\\ 
& P2022-07/08 & PKS 1424-418 & FSRQ  & 1.52 & 31.7 & LAT & \xmark  & \\ 
& P2022-12 & 3FHL J0543.9-5532 & HBL  & 0.273 & 27.5 & VHE & \cmark  & \\
\hline
\multirow{2}{*}{2023}& P2023-02 & PKS 0402-362 & FSRQ  & 1.42 & 9.9 & LAT & \xmark  & \\ 
& P2023-05 & PKS 1424-418 & FSRQ  & 1.52 & 3.5 & LAT & \xmark  & \\ 
\enddata

\end{deluxetable}

\clearpage
\bibliography{skeleton.bbl}
\bibliographystyle{aasjournal}

\clearpage
\section*{Full Authors List: H.E.S.S. Collaboration}
\include{authors_ALL_ICRC}

%
%
%

\end{document}

%% file: authors_ALL_ICRC.tex
\scriptsize
\noindent
F.~Aharonian$^{1,2,3}$, 
F.~Ait~Benkhali$^{4}$, 
A.~Alkan$^{5}$, 
J.~Aschersleben$^{6}$, 
H.~Ashkar$^{7}$, 
M.~Backes$^{8,9}$, 
A.~Baktash$^{10}$, 
V.~Barbosa~Martins$^{11}$, 
A.~Barnacka$^{12}$, 
J.~Barnard$^{13}$
% ...

\medskip

\noindent
$^{1}$Dublin Institute for Advanced Studies, 31 Fitzwilliam Place, Dublin 2, Ireland\\
$^{2}$Max-Planck-Institut f\"ur Kernphysik, P.O. Box 103980, D 69029 Heidelberg, Germany\\
$^{3}$Yerevan State University,  1 Alek Manukyan St, Yerevan 0025, Armenia\\
$^{4}$Landessternwarte, Universit\"at Heidelberg, K\"onigstuhl, D 69117 Heidelberg, Germany\\
$^{5}$IRFU, CEA, Universit\'e Paris-Saclay, F-91191 Gif-sur-Yvette, France\\
$^{6}$Kapteyn Astronomical Institute, University of Groningen, Landleven 12, 9747 AD Groningen, The Netherlands\\
$^{7}$Laboratoire Leprince-Ringuet, École Polytechnique, CNRS, Institut Polytechnique de Paris, F-91128 Palaiseau, France\\
$^{8}$University of Namibia, Department of Physics, Private Bag 13301, Windhoek 10005, Namibia\\
$^{9}$Centre for Space Research, North-West University, Potchefstroom 2520, South Africa\\
$^{10}$Universit\"at Hamburg, Institut f\"ur Experimentalphysik, Luruper Chaussee 149, D 22761 Hamburg, Germany\\
$^{11}$Deutsches Elektronen-Synchrotron DESY, Platanenallee 6, 15738 Zeuthen, Germany\\
$^{12}$Obserwatorium Astronomiczne, Uniwersytet Jagiello{\'n}ski, ul. Orla 171, 30-244 Krak{\'o}w, Poland\\
$^{13}$Department of Physics, University of the Free State,  PO Box 339, Bloemfontein 9300, South Africa\\
% ...
